\documentclass{article}
\usepackage{color}
\usepackage{graphicx}
\usepackage{hyperref}
\usepackage{pstricks,amssymb}
\usepackage{fancyhdr}
\usepackage[raggedright]{titlesec}
\begin{document}
\title{Quantum view of Mass\\
\vspace{.3cm}
\large{ R. Ramachandran}\\
\vspace{0.3cm}
\normalsize{
Institute of Mathematical Sciences\footnote{Affiliated as retired Faculty; Communication address: Flat 12, Khagol Society, Panchvati, Pashan, Pune 411 008, India}, Taramani \\
Chennai, TN 600113, India.\\
{\small rr@imsc.res.in, rr\_1940@yahoo.co.in.unipune.ac.in}\\
\vspace{0.3cm}
(Submitted 25-08-2012, Revised 08-09-2012)}\\
\rule[0.1cm]{16cm}{0.02cm} \\
\textbf{Abstract}\\
\flushleft \normalsize {The classical view of mass is that it quantifies the amount of substance and is a kinematical parameter. All matter has an attribute of mass and is a conserved quantity in any interaction. With the advent of Special Relativity, mass became no longer a conserved quantity, since Energy and Momenta had the status of conserved variables. Nevertheless, $\sqrt{(E^2-{\bf p}^2c^2)}= mc^2$ gives a Poincare invariant measure that can be associated as the mass, an useful attribute of the body or system. In the quantum regime mass becomes truly dynamical. Higgs field is said to provide mass for all species of elementary constituents – as widely popularized by the media in connection with the recent (most likely) discovery of Higgs boson at CERN.  However, we emphasize that the most abundant component of matter – Nucleons - derive their mass largely (95\%) as a consequence of quantum effects of (color gluonic QCD) radiation. Further, interestingly this arises out of literally nothing, save the QCD scale, determined experimentally through a self consistent perturbative analysis of nucleon structure, as the sole input. } \\
\rule[0.1cm]{16cm}{0.02cm} 
}
\date{}
\maketitle
\thispagestyle{fancy} 
\renewcommand{\headrulewidth}{0.4pt}
\renewcommand{\footrulewidth}{0.4pt}
\section{Introduction}
Higgs particle discovery \cite{cern1},\cite{cern2},\cite{toinewsrep} has received much coverage and a perception that the so called ‘God Particle’ is responsible to give masses to all particles that make up the fundamental building blocks is prevalent. The object of this note to give a more appropriate perspective and provide non experts, particularly Physics Teachers and Students, a deeper view on what constitutes mass and what has been understood so far\footnote{An excellent review is provided by F Wilczek \cite{wilczek}; supplement to 2011 Solvay conference, amplifies the content of this note}. 

In classical physics, mass is a kinematical attribute of all matter. It is a measure of the quantity of matter and is perceived through two laws, both attributed to Newton. Force causes matter to accelerate and the proportionality constant is termed its inertial mass. Matter is also source of gravitational field it carries with it. This field falls off in intensity as Inverse Square of the distance from the source and the proportionate constant here is its gravitational mass. Galileo's famous experiment (at the leaning tower of Pisa?) and many modern equivalents demonstrate identity between the two definitions of mass and this implies a notion of universality of all bodies under gravitation. In classical regime the mass is a passive kinematical parameter and is conserved in any interaction. As we move to relativistic regime, we find that it is not mass that is conserved, but the Momentum (vector) {\bf p} and Energy E. There is, however an invariant mass for every body or system which is given by  $\sqrt{(E^2-{\bf p}^2 c^2}/c^2$. Even this (Poincare invariant) mass is not conserved in any interaction, since mass of the system can be released as energy, heralding the celebrated relationship $E = mc^2$. In the terminology of Nuclear Physics, the mass defect shows up as the binding energy of nucleons in nuclei. Lighter nuclei such as Hydrogen, Helium and Carbon fuse to form tighter bound nucleus releasing the difference in mass as thermonuclear energies in a fusion reaction; and heavy nuclei, such as Uranium and Plutonium can be induced to undergo fission into medium heavy nuclei, releasing useful atomic energy, making in the process the iron region nuclei with highest binding energy per nucleon. 

In quantum regime, we see that mass, whatever it may be, is dynamically generated. The notion of mass defect is an indication that the mass of a system, say an atom or nucleus is made up by a combination of the intrinsic mass of the constituents suitably dressed by interactions. The system may have a higher or lower mass than the sum total of constituents, making it either a resonant state or a bound state. For example, the energy spectrum of an atom is a consequence of the interaction of the constituents. Electromagnetic interaction between positively charged nucleus and negatively charged electrons results allowed energy levels in the atomic spectrum. To begin with, we have Schroedinger equation in Quantum mechanical description of an atom, say Hydrogen, give observed values of its spectra. This can be further improved and made fully relativistic in the language of Quantum Field Theory. Relevant field theory to deal with atomic (and molecular) spectra is Quantum Electrodynamics (QED), which comes endowed with Gauge symmetry. Gauge symmetry is a formal way of implementing a notion that the Electric and Magnetic field, that enters in the Lorentz Force law ${\bf F} = q({\bf E} + {\bf v} \times {\bf B})$ is expressible in terms of scalar and vector potentials (${\bf E} = -{\bf \nabla} \phi - {\partial {\bf A} \over \partial t}$; ${\bf B} = {\bf \nabla} \times {\bf A}$). There is a freedom in the choice, since ${\bf A} \to {\bf A}' = {\bf A} + {\bf \nabla} \chi({\bf x}, t)$ and $\phi \to \phi' -{\partial \chi({\bf x}, t) \over \partial t}$ leaves ${\bf E}$ and ${\bf B}$ invariant. Gauge theories formulated in terms of potential functions  instead of field functions, necessary in quantum description (since Aharonov - Bohm \cite{AB} effect shows that the quantum electrons passing through magnetic field free region indeed detects change in the interference pattern, making potentials more fundamental than field strengths),  comes endowed with a symmetry so generated. Since there is one function that characterises this symmetry, mathematically this is represented by a unitary unimodular group $U(1)$. We are able to achieve highly precise computations of the Energy levels (or masses) and transition rates, thanks to  very reliable perturbation techniques, developed in the later half of the last century.  

Proceeding further, nucleons in the nuclei are bound together by strong nuclear forces and the nucleons are indeed made up of quarks, bound by interactions mediated by gluons. Quarks and Leptons (electrons and the siblings) are the building blocks of all  matter in the Standard Model. Like QED, the Standard Model is also a Gauge field theory with an underlying local \footnote{The term `local' implies the symmetry transformation parametrised by $\chi({\bf x},t)$ is a spacetime dependent function. If $\chi$ is a constant value independent of space and time, we will have a `global' symmetry, such as flavour Isospin} symmetry described by a symmetry group. While QED with symmetry group $U(1)$, [using relativistic four dimensional potentials $A_\mu$ that combines ${\bf A}$ and $\phi$, a one parameter change $A_\mu \to A'_\mu = A_\mu - ie\partial_\mu \chi $ leaves the field strength $F_{\mu \nu} =\partial_\mu A_\nu - \partial_\nu A_\mu$ unchanged] admits one (well known electric) charge and one gauge field whose quanta are photons, the Standard model has underlying symmetry group as $SU(3) \times SU(2) \times U(1)$ admits 3 coupling `constants', 8+3+1 parameter symmetry transformations and has force fields as due to gauge bosons (8 gluons, 3 weak bosons and photon). They govern strong, weak and electromagnetic interactions of all basic constituents. An important difference is that while $U(1)$ of QED is commuting symmetry group (where two symmetry operations, one following the other in either order gives same result), the gauge group of Standard Model has non commuting components $SU(2)$ for weak interactions and $SU(3)$ of QCD for strong interactions. Here symmetry operations are represented by unitary unimodular $2 \times 2$ and $3 \times 3$ matrices. These are similar in character to rotations in space, which we know, in three or more dimensions, to be non-commuting. We may refer them as operations in some internal weak isospin and color space.    
\section{Massless Start }
Theoretical description of basic interaction employs the tools of Relativistic Quantum Field Theory in the form of Gauge Theory. There are three pillars on which it stands and each of which needs, to begin with, massless fields as basic input. 

\subsection{Scale Invariance and Renormalization} 
 We need our theories to be so that all observables yield finite values. It is necessary to prevent divergences, if any, from appearing in any measureable variable. This technical requirement is achieved by the process of Regularisation and Renormalization and this program is successful on account of the theory possessing scale invariance. In a scale invariant theory it is possible to promote the coupling constants, such as the `fine structure constant' $\alpha$ here, into scale dependent parameters. The constant $\alpha = e^2/4\pi \hbar c$  becomes $\alpha(Q^2)$ and it measures the coupling strength or charge at different scales.

  It is said that the `Vacuum polarization' causes the bare charge to be screened, making charge depend on the scale of the probe used. A simple way to understand renormalisation is to note  that in the quantum regime, `vacuum' is anything but simple, since it can be thought of as all types of particle and  antiparticle pairs to be continually created and annihilated perpetually, making it a polarisable medium or an effective dielectric. Just as effective charge in a dielectric medium gets reduced by the dielectric constant of the medium, a negatively charged electron with bare charge $e_0$ will polarise the nearby region of the `vacuum' and consequently the measured charge will be the screened value. $e(Q^2)$ will be the effective charge when we approach it with a probe that causes a momentum transfer $Q^2$. Larger the value of $Q^2$, closer we approach it and lesser the screening. The value $e = 1.6 \times 10^{-19}$ Coulomb or $\alpha = 1/137$ is indeed the long range Thomson limit, when $Q^2=0$. 
  For all this to make sense, the theory must possess an intrinsic scale invariance. A closely related symmetry is the angle preserving conformal invariance. Since, angle is ratio between two lengths, it does not change under scale transformation that varies all lengths in the same way. 
It is often convenient to use a set of units, such that $\hbar = 1 = c $ and in such units dimension of mass is just the inverse of the dimension of length. Recall the Compton wavelength $\lambda$ associated with mass $m$ is given by $\lambda = \hbar /mc$. If the theory has mass parameter, it possesses an intrinsic length; obviously such a theory can not be scale invariant. Thus basic ingredients in a renormalizable theory have necessarily to be massless. Presence of mass will imply scale violation.
\begin{figure}[!h]
\centering
\includegraphics[width=3in,height=2in]{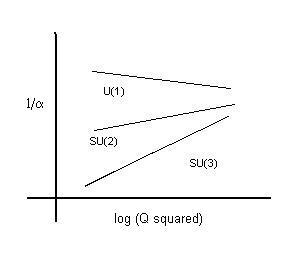}
\caption{\small ${1 / \alpha (Q^2)}$ vs $log \ Q^2$ }
\label{fig:af}
\end{figure}

\subsection{Chiral Fermions}
Among the basic interactions are weak interactions, responsible for radioactivity. As early as 1957, we knew these to be parity violating. In order that this is so, we need to differentiate between the left handed and right handed states of fermions. In the Standard Model the left helicity states of quarks and leptons are doublets (in weak isospin, not to be confused with the more widely known flavor isospin),
$Q_L = \pmatrix{u_L \cr d_L}$, $\psi_L = \pmatrix{\nu \cr e_L^{-}}$ 
and the right helicity states of fermions $u_R$, $d_R$ and $e_R$ are singlets. Neutrino occurs in left helicity state only.   If we view the fermion from a frame that moves faster than the particle (which is possible only if the particle has mass and travels with speed $v < c$) we will find in that frame, the helicity of the particle is reversed. Thus a chiral fermion, which is forced on us by the parity violation, is not compatible if the fermion has a mass. We require our fermions to be massless in order that they are viewed as chiral fermions. Fermionic matter consists of three copies (or three generations) of the above set, which is again forced on us as a need to accommodate a baryon asymmetric universe, which is populated mostly with nucleons with negligible fraction $(~10^{-10})$ of anti-nucleons. That is another story. Neutrinoes  also seem to mix and oscillate, which is possible when they have a tiny mass. That is yet another story.  

\subsection{Gauge interaction and massless bosons}
The Standard Model describes interactions governed by the gauge theory with symmetry group $SU(3)_C \times SU(2)_L \times U(1)_Y$. The subscripts $C, L, Y$ refer to color (strong/chromodynamic), weak Left helicity isospin and a weak hypercharge $Y$ respectively. Correspondingly there are gauge bosons; octet of gluons, electroweak bosons $W^i$, i =1, 2, 3 and $W^0$. $W^3$and $W^0$ combine to form two orthogonal combinations, of which one is the familiar electromagnetic photon $\gamma $, call it $A$ and the other neutral weak boson $Z$. Together with $W^{1{\pm} i2}=W^{\pm}$, we have weak intermediate bosons that mediate both neutral and charge changing weak interactions. 

The Lagrangian density of the Standard model is given as: 
\begin{eqnarray}
\nonumber {\mathcal{L}} &=& -{1\over 4} {G^a_{\mu \nu}G^{\mu \nu}_a} -{1\over 4} {W^i_{\mu \nu}W^{\mu \nu}_i}
-{1\over 4} {W^0_{\mu \nu}W^{\mu \nu}_0} 
\\
\nonumber &&-\overline{Q_L}\gamma^\mu (\partial_\mu - ig_3\lambda_a G^a_\mu -ig_2{\tau_i \over 2}W^i_\mu -{1 \over 6} ig_1W^0_\mu)Q_L 
\\
\nonumber &&+ \overline{u_R}\gamma^\mu (\partial_\mu -\lambda_a G^a_\mu -{2 \over 3}ig_1 W^0_\mu)u_R 
 + \overline{d_R}\gamma^\mu (\partial_\mu -\lambda_a G^a_\mu +{1 \over 3}ig_1W^0_\mu)d_R 
 \\
\nonumber &&+\overline{\psi_L}\gamma^\mu (\partial_\mu -ig_2{\tau_i \over 2}W^i_\mu +{1 \over 2} ig_1W^0_\mu)\psi_L +\overline{e_R}(\partial_\mu + ig_1W^0_\mu)e_R
\end{eqnarray} 
 
\noindent where $G^a_{\mu \nu} = \partial_\mu G^a_\nu -\partial_\nu G^a_\mu + f_{abc}G^b_\mu G^c_\nu$,  with $a,b,c$ taking values $1,2..8$, $f_{abc}$ being the structure constants of $SU(3)$;  $W^i_{\mu \nu} = \partial_\mu W^i_\nu -\partial_\nu W^i_\mu + \epsilon_{ijk}W^j_\mu W^k_\nu$, here  $i,j,k$ assume values $1,2$ and $3$, $\epsilon_{ijk}$ are structure constants for $SU(2)$; and $W^0_{\mu \nu} = \partial_\mu W^0_\nu - \partial_\nu W^0_\mu$. $\lambda_a$ is the set of eight $3\times 3$ Gellman matrices related to the generators of $SU(3)$ symmetry just as $\tau_i$ are three $2 \times 2$ Pauli matrices that occur in the generators of $SU(2)$ symmetry.

$G^a_\mu$ is the octet vector field of of Gluons of QCD. $W^\pm_\mu = W^1_\mu \pm iW^2_\mu$ are the charged intermediate vector bosons that couple to charge changing weak currents, responsible for radioactivity ($\beta$ decays); and the vector bosons $W_\mu^3$  and $W^0_\mu$ combine to become the conventional photon field $A_\mu (= {g_1 W_\mu^3+g_2 W^0_\mu \over \sqrt{g_1^2+g_2^2}})$ and the orthogonal neutral gauge boson $Z_\mu$ that is responsible for neutral weak current interactions. The coefficients of $g_1$  in the equation above reflect the weak hypercharge $Y$ of the fermion field in that term. The coupling parameters $g_i$, as discussed in the preceding section on account of the renormalization process, get promoted into scale dependent functions $g_i(Q^2)$, where  $Q^2$ is the square of the momentum transfer used to probe and their evolution as a function of $Q^2$ depends on what is known as the beta function [$\partial g/\partial log Q^2 = \beta(g)$],  of the respective symmetry group. A characteristic feature of these functions is that they make $g_3$ and $g_2$ logarithmically decrease as  $Q^2$ increases, [while in contrast we have logrithmically increasing property for $g_1$] reflecting thus the anti-screening of the non-abelian charges.  At extremely short distances, which need high momentum transfers and hence high energies to probe, the coupling is asymptotically vanishing. See the sketch in fig 1. Quarks color interactions are then small, amenable to perturbation treatment. Quark interactions are said to enjoy asymptotic freedom \cite{GW}, \cite{politzer} Deep inelastic scattering (high energy, high momentum transfer) by $e$ or $\mu$ off nucleon targets revealed that quarks inside the nucleons can be regarded as free and non interacting! 
  
	Notice that there is no term quadratic in the gauge fields, such as $G_\mu^a G_a^\mu$, $W_\mu^i W_i^\mu $or $W^0_\mu W_0^\mu$, signifying that gauges bosons are like massless photons. There is no way to introduce a gauge preserving mass term. However, if the intermediate vector boson is massless, this will make the weak radioactivity  a long ranged effect like electromagnetism! The mechanism to give masses to gauge bosons (without ruining the gauge symmetry), so that weak interactions remain short ranged is the celebrated Higgs mechanism. It  achieves two outcomes. It makes the symmetry hidden (also referred to as spontaneously broken) in a way that the solution of the theory reflects a lesser symmetry (in our case $SU(3)_C  \times U(1)_{em}$ than that of the underlying Lagrangian. The gauge bosons associated with the so called hidden symmetries, for us $W^\pm$ and $Z$, acquire mass. Further through the coupling the Higgs field has with all matter fermions, it also generates their masses. The minimum Higgs scheme calls for a new complex (weak isospin ${1 \over 2}$) doublet scalar field $\Phi =\pmatrix{\phi^+ \cr \phi^0}$, which is a color singlet and carries one unit of weak hypercharge $Y$. Together with its hermitean conjugate, we now have 4 scalar fields added through the Higgs phenomenon, with Lagrangian density (note wrong sign of $\Phi^\dagger \Phi$ ‘mass’ term)as
$$
\mathcal{L}_{Higgs} = - [D_\mu \Phi^\dagger D^\mu \Phi - \mu^2 (\Phi^\dagger \Phi) +\lambda (\Phi^\dagger \Phi)^2 ]
$$
where the covariant derivative term
$$
D_\mu \Phi = (\partial_\mu -ig_2{\tau_i \over 2} W^i_\mu  -{1 \over 2}ig_1 W^0_\mu )\Phi
$$
\begin{figure}[!h]
\centering
\includegraphics[width=3in,height=2in]{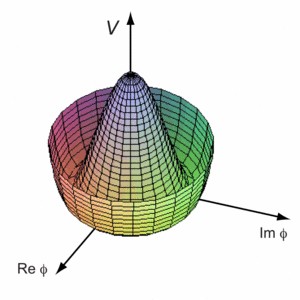}
\caption{\small $V(\Phi)$ vs $\Phi$ illustrating degerate vacuum states}
\label{fig:mexhat}
\end{figure}
The shape of $V(\Phi) = -\mu^2 (\Phi^\dagger \Phi)+ \lambda(\Phi^\dagger \Phi)^2$, resembles a mexican hat as shown in the figure 2. On extremizing $V(\Phi)$, we find $\langle \Phi \rangle =0$ as an unstable maxima in the central peak and a degenerate set of minima at the bottom of the valley, each of which can be the vacuum state, all characterized by a nonzero value of the Higgs field. We may choose the vacuum state to be one of them, say, given by   
$$
\langle \Phi\rangle = {1 \over \sqrt{2}} \pmatrix{0 \cr v}, \mbox{where}\  v = {\mu \over \sqrt{\lambda}}.
$$

It is easily verified that the unbroken generator is $I_L^3+Y/2 $, that links the set of vacuum states  and we associate it with the electromagnetism. Presence of non-vanishing $v$ gives masses to $W^\pm$ and $Z$ that are indeed observed at CERN with mass values  80 GeV and 91 GeV respectively. By a field redefinition we can demonstrate that three of the four Higgs fields metamorphose into the longitudinal modes of $W^\pm$ and $Z$ bosons (now that these gauge bosons are massive they should  have all three spin polarizations as against there being only two transverse polarizations for (massless) radiation), leaving one surviving mode, the recently discovered \cite{cern1},\cite{cern2} Higgs boson at 125 Gev.
	
	Higgs coupling with the fermions (Yukawa interaction) provides masses for all fermions, such as quarks and leptons, the value of the mass being proportionate to the coupling parameter.  Yukawa terms in the Lagrangian that give fermions their masses as well as their interactions is given by
$$
\mathcal{L}_{yukawa} = h_u\overline{Q_L}\Phi u_R + h_d \Phi d_R  + h_e \overline{\psi_L}\Phi e_R + h.c.. 
$$
$h_u$, $h_d$ and $h_e$ are free parameters and are proportional to the relevant quarks and electron masses. Masses of all fermion constituents and vector bosons (the quanta that mediates  forces) derive their mass values as a consequence Higgs phenomenon. This is the sense in which it is claimed that Higgs field, that pervades all space generates mass for the constituents in the universe.
	
	We wish to point out that this is a bit of an exaggeration, given the fact the mass generated by the phenomenon yields very tiny values (2.15, 4.70 and 0.51 MeV) for $u$, $d$ quarks and electron, which form almost all stable matter found in the universe. The bulk of mass for nucleons, in fact, arises from a different beautiful phenomenon and it is remarkable that this is a consequence again of the quantum principle. Recalling that spin (half integral angular momentum) of fermion has no classical analogue, we may assert that mass and spin are quantum attributes with little underpinnings in classical Physics. 

\section{Mass out of Nothing}
We may suspend for a while the Higgs phenomena and deal with just strong and electromagnetic regime. The dynamics is governed by $SU(3)_C \times U(1)_{em}$ gauge theory that survives electroweak symmetry breaking. While electrodynamic forces govern the atomic structure of all elements and thereby all of chemistry, QCD is responsible to give us the nucleons and mesons as color neutral bound states of quarks with  gluons as carriers of chromodynamic forces. Further, the residual (van der Waals like) interactions mimic the strong short range nuclear forces among nucleons and mesons build up the various nuclei, much like molecules are built out of electrically neutral atoms. The non-abelian gauge group is bestowed with asymptotic freedom (or vanishing coupling at very high frequencies or very short distances) and reciprocally confinement of color. Quarks and Gluons, that carry color quantum number are not to be seen as asymptotic states and are permanently confined within the color singlet modes. Mesons and Baryons as solutions of the dynamics constitute the spectrum of states. Their masses and the transition rates among them can be computed in QCD, just like QED provide {\it ab initio} atomic and molecular spectroscopy. Extreme precision in atomic spectroscopy and optics have been possible as a consequence of the development of high precision perturbative computation, since the small dimensionless coupling parameter $\alpha = 1/137$  renders reliability and order by order convergence of the computed quantities. In nuclear physics we do not have a small parameter to help us. However, in the underlying strong interaction, which we now recognize as emerging from QCD, it is possible to invoke perturbative QCD for a short distance (high $Q^2$) probe and use it to find both the scaling and quantitative scaling violation in deep inelastic scattering of leptons off protons and neutrons in the nuclei. This theory, however, is neither useful at predicting the low energy spectrum of baryons and mesons, nor determine the wave-functions of quarks in the hadrons. We need turn to non perturbative attempts to understand these features of QCD. 

Lattice gauge theory reconstructs the theory on a space-time made up of lattice, such that as the lattice spacing is reduced and vanishes, the continuum theory is recovered. Methods of statistical mechanics are used to compute the various correlation functions and extract values for physical observables, such as masses of the bound states and resonant states, transition rates etc. of the theory, given just a few parameters that define the theory. We refer the reader to several review articles available (see \cite{pdg}) and give here just an overview of what goes into the theory and the outcome thereof. 

First, if there is no Higgs mechanism and no mass scales in the theory, how do we generate mass for the observed state? While classically the theory is scale invariant and hence has no mass parameter in the theory, when we deal the problem in quantum regime, a scale gets introduced as a process of regularization and renormalization. The coupling constant becomes a scale dependent parameter (Sidney Coleman called it a dimensional permutation \cite{coleman}). Scale invariance now implies that as scale is changed there is a definite way all measured observables change. It also serves to shield intrinsic divergences, if any, in the theory to remain hidden in unobservable parameters of the theory. This provides us with a prescription to compute all measureable quantities in terms of a few parameters of the theory. QCD is defined with an intrinsic reference scale $\Lambda_{QCD}$ at around 100 MeV, which we determine experimentally from the perturbative analysis of the deep inelastic scattering of leptons off nucleon targets \cite{pdg}. 

The ingredients of theory is that we have a $SU(3)$ color gauge theory endowed with a fermion content made up of three generations of quarks. After the Higgs phenomenon we have quarks acquiring mass and phenomenological observation is that there are three light quarks and three heavy quarks. Of these $u$ and $d$ quarks are very light, $c, b, t$ quarks are very heavy and $s$ quark in the same order as $\Lambda_{QCD}$. We may begin with a toy model (Wilczek calls it QCD lite), setting all light quarks $u, d$ and $s$ massless and $c, b$ and $t$ infinite. The heavy flavours naturally decouple; and the three light massless quarks in the computation should give us a flavour $SU(3)$ spectra. Particle phenomenology of hadrons reflect an approximate flavour $SU(3)$ (with isospin $I$ and  hypercharge $Y(=B+S)$) symmetry, known to consist of a pseudoscalar meson octet ($\pi^\pm$, $\pi^0$,
 $K^\pm$, $K^0$, $\overline{K^0}$, and  $\eta$), a vector meson nonet ($\rho^\pm$, $\rho^0$, $K^{*\pm}$, $ K^{*0}$, $ \overline{K^{*0}}$, $\omega$ and $\phi)$, a baryon octet ($p$, $n$, $\Lambda$, $\Sigma^+$, $\Sigma^0$, $\Sigma^-$,$\Xi^0$ and $\Xi^-$) and an excited baryon decimet ($\Delta^{++}$, $\Delta^+$, $\Delta^0$, $\Delta^-$, $\Sigma^{*+}$, $\Sigma^{*0}$, $\Sigma^{*-}$, $\Xi^{*0}$, $\Xi^{*-}$ and $\Omega^-$) as the prominent low energy spectra. Lattice gauge theory computations are able to quantitatively postdict this spectra. This computation has no input parameters, save the notion that $g_3(Q^2)$ depends on QCD scale $\Lambda_{QCD} $, which is obtained perturbatively from studying scaling violations of proton structure functions in deep inelastic scattering. In these experiments one uses weak and electromagnetic probes $(e, \mu, \nu)$ to get the hadronic structure functions, whose $Q^2$ dependence gives us the scale of QCD. With only $\Lambda_{QCD}$ as input, we make a statistical analysis of the system in a lattice framework of QCD. They yield a value $M_{p,n}\sim 890 $ MeV, thus almost accounting for $95\%$ of its mass as arising out of mass-less quark gluon radiation reaction. 

\section{Realistic hadron spectra from Lattice QCD}  

Lattice Gauge Theory aims to study Quantum Chromo Dynamics on a sufficiently large space-time (with periodic boundary conditions in all directions) regarded as a 4 dimensional grid of volume $L^4$ with lattice spacing of length $a$. A space time point $x_μ$ is specified by four integers through $x_\mu = n_\mu a$ and the limit $a \to 0$ and $L \to \infty$  lets us pass to continuum theory.  Quark degrees of freedom $q_f(x)$, $f = u,d, ..$ reside on the lattice points and the gauge fields, gluons and photons, within $U_{\hat{\mu}}(x)= exp(i\int_x^{x+\hat{\mu}} G_{\hat{\mu}}(x')dx')$  on the links (numbering 8 for each site) that join a pair of neighbouring lattice sites.   One then defines the partition function as the integral over all field variables of the Standard Model action of 
 Gluons and Fermions;
 
$$ 
S= S_G + S_F. $$ $$
Z=\int DU D\psi D\overline{\psi}\  exp(-S[U,\psi,\overline{\psi}])
$$

Statistical averaging of all possible configurations of the fields on the lattice allows us to simulate QCD and compute sampling of various field configurations on it. From these it is possible to extract experimentally measureable quantities. Powerful computational algorithms back up the effort to extract from it the outcome of the particle spectra and various transition amplitudes. 


\begin{figure}[!h]
\centering
\includegraphics[width=3in,height=2.5in]{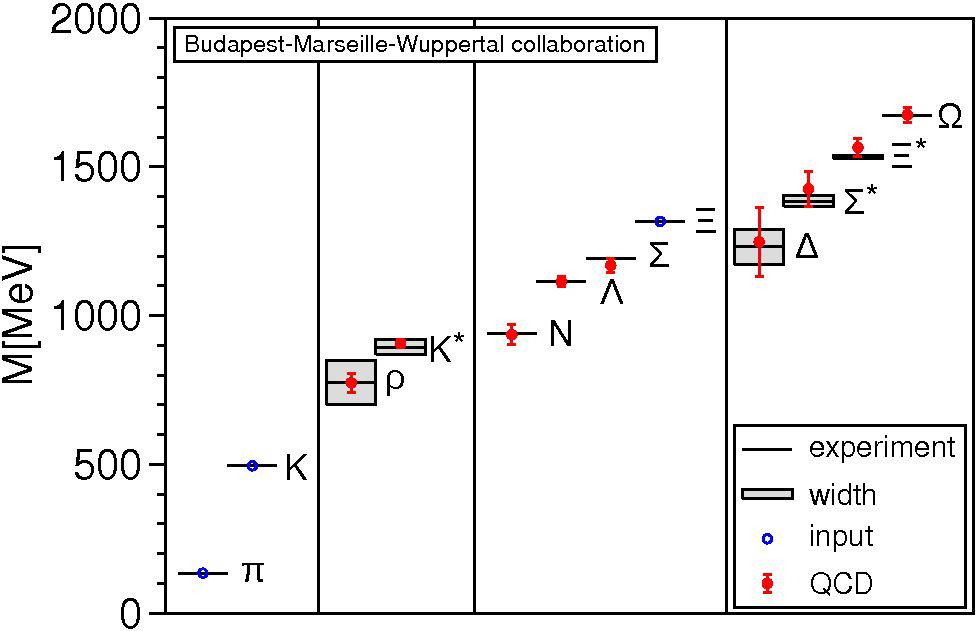}
\caption{\small Spectrum of low energy spectra, computed from first principles in QCD; source \cite{durr} }
\label{fig:spectra}
\end{figure}

We saw in the preceding section computations with exact chiral invariance (with massless quarks) to obtain nucleon mass as 890 MeV. Next step is to let  the parameters for quark masses $m_u$ = $m_d$ and $m_s$ free. In a Full QCD computation recently reported, BMW Collaboration, \cite{durr} used state of the art lattices with L/a = 64 and thus the space time has $N = 64^4 = 16,777,216$ sites. Computation involved matrices of dimension $12N \times 12N $ and storing about $4 \times 10^{16}$ complex numbers. Adjoining figure illustrates the results. With pion ($\pi$), kaon ($K$) and cascade baryon ($\Xi$) masses as input values, we get the values of $\rho$, $K^*$, $N$, $\Lambda$, $\Sigma$, $\Delta$, $\Sigma^*$, $\Xi^*$ and $\Omega$. We have a remarkable agreement in the description of the observed set of pseudo scalar mesons, vector mesons, baryons and excited baryon states. Nucleon is, as expected, at 940 MeV.  
 
Lattice characterization of QCD, should not be seen as an approximation to continuum space-time, but (generically) an unavoidable interim part in the definition of the theory. The procedure is intrinsically gauge invariant since it deals with gauge invariant content all the time, unlike in a perturbative treatment, where a choice of a gauge has to be made and care must be exercised to ensure that the final outcome is gauge invariant. This Lagrangian - regulated renormalized extrapolation - respects confinement of color, chiral symmetry and scale invariance limits adequately. We are able to address many features of QCD in the low energy regime of hadron physics that are available for experimental study; such as decay constants $f_\pi$,  $f_K$; semileptonic form factors that are appropriate for computing $B \to Dl \nu , Kl \nu, \pi l \nu;\ l = e,\mu $ etc. To begin with one computes in the so called quenched approximation, in which quark degrees of freedom are ignored in order to keep the `cost' of computation in terms of available computing resources kept within manageable level. As tera and penta flop speed in computing get developed more ambitious project of `full' QCD are possible as a major global collaborative endeavour. 
We will paraphrase Wilczek\cite{wilczek} in identifying the conceptual roots that shaped the outcome as represented in the fig. 2. Special theory of Relativity appears to demand that the interactions are local; the local interaction bring in fields with arbitrarily large frequencies (energy) and short wavelengths (high momenta) that may cause divergence that will render the calculations unreliable. Non abelian gauge theories avoid it, by virtue of the property of asymptotic freedom that weakens the coupling of the dangerous modes. This happy result occurs only for the gauge invariant minimal couplings as are considered in these exercises. 

QCD, a gauge theory based on gauge group $SU(3)$ color triplet quarks and color octet gluons, both degrees of freedom remaining confined in gauge singlet hadrons is highly constrained, supporting very few free parameters. A mass parameter for each flavor quark (together with flavor mixing angles of Cabibbo - Kobayashi – Masakawa matrix) and just an overall coupling strength is all that one is allowed.  Since asymptotic single quark states are never seen, the mass parameters of quarks are to be seen as just inputs that figure in getting the masses of hadrons. The coupling $\alpha_s(Q^2)=g_3 (Q^2 )/4\pi \hbar c $  is large when Q is less than or of order $\Lambda_{QCD}$ and fluctuations in gluon dominates the dynamics. Bulk of the nucleon mass, we may presume, thus gets built on a tiny chiral symmetry breaking mass of $u$ and $d$ quarks by the gluon dressings carrying most of energy associated with the state. 
This is reminiscent of what was indeed an old speculation of Lorentz as the origin of electron mass. He associated rest-mass/energy of the electron with the energy in the form of Electric field residing in the space, $1/(2\epsilon_0) \int d^3 x E^2 (x)$, sort of radiation reaction on the motion of the electron. For a point charge this will be indeed divergent, but is finite for an electron with a distributed size of range $\lambda =\hbar/mc$, its Compton wavelength. We may use this to fix the radius of electron (which turns out be of order $\alpha \lambda$). While this is not anymore regarded as the origin of electron mass, (now that $\alpha$ is no longer a fixed constant and the Higgs coupling rather than finite size of electron as dictating it) we find that the hadron masses seem to possess some features of gluonic radiation reaction as generating bulk of the mass, in a somewhat similar picture as that of Lorentz. It is remarkable that Lattice framework of QCD provides dependable {\it ab initio} prediction, thanks to high speed computing resources and very smart dedicated algorithms available now for such a computation. 
\section{Summary}
While classically mass is an extrinsic kinematic parameter that signifies the amount of matter, quantum regime makes {\em mass} a dynamical input. This feature for mass has two somewhat different origins. First, we observe that Quantum Chromo Dynamics (QCD), that governs interaction among quarks and gluons, is responsible for the mass to primary nucleons, the most abundant source of visible matter in the universe. Next, apart from the $u$ and $d$ quarks (and the leptons $\nu_e$ and $e$) we need at least two more generation of quarks and leptons to complete the matter content. All of them were abundant and in a dynamical equilibrium at the very early stages of the universe, but now most (except quarks and leptons of the first generation) are only seen as short lived intermediate particles. These as well as the weak interaction-mediating bosons  $W^\pm$ and $Z$ get their mass as a result of the coupling with the Higgs scalar field. We emphasize that both features point to the notion that {\em mass} is essentially a quantum consequence, which has no classical analogue.    

\end{document}